\documentclass[a4paper,11pt]{article}
\usepackage{jinstpub} 
\usepackage{lineno}

\usepackage[figuresright]{rotating}

\usepackage{mathtools}
\usepackage{times}
\usepackage{latexsym}
\usepackage{graphicx}
\usepackage{booktabs}
\usepackage{xcolor}
\usepackage{verbatim}

\usepackage[T1]{fontenc}
\usepackage[utf8]{inputenc}
\usepackage{microtype}
\usepackage{float}

\usepackage{hyperref}
\usepackage{cleveref} 
\usepackage{pdfpages}
\usepackage{subfig}

\usepackage{epstopdf}
\usepackage{subcaption}
\usepackage[export]{adjustbox}
\usepackage{changepage}
\usepackage{graphicx}
\usepackage{subcaption}
\usepackage[section]{placeins}

\usepackage{algorithm2e}

\definecolor{myMaroon}{RGB}{128, 0, 0}
\definecolor{mySalmon}{RGB}{250, 128, 114}
\definecolor{myGreen}{RGB}{75, 150, 75}

\newcommand{\citee}[1]{\textbf{ToDo Ref}}

\title{\boldmath A decomposition algorithm for streak camera data}

\author[a,1]{Kaan Oguzhan,\note{Corresponding author.}}
\author[a]{Lucas Ranc,}
\author[b,2]{Livio Verra,
\note{Now at: INFN Laboratori Nazionali di Frascati, Frascati, Italy}}
\author[a]{Allen Caldwell}
\affiliation[a]{Max Planck Institute for Physics, Munich, Germany}
\emailAdd{kaan.oguzhan@tum.de}
\emailAdd{ranc@mpp.mpg.de}
\emailAdd{caldwell@mpp.mpg.de}
\affiliation[b]{CERN,Geneva, Switzerland}
\emailAdd{livio.verra@cern.ch}

\begin{document}



\abstract{
  We describe a novel reconstruction algorithm for time-resolved images obtained using a streak camera. This algorithm operates by decomposing a recorded image into a set of individual photoelectron-induced signals, thereby providing a powerful method for streak camera image reconstruction. This deconstruction allows for a standard statistical analysis of the resulting image. We demonstrate the effectiveness of this technique by analyzing the temporal spacing between the emitted fs-long laser pulse and its succeeding first, second, and third reflections within a thick glass captured by a streak image.

}\label{sec:abstract}

\keywords{Beam-line instrumentation, Data processing methods, Analysis and statistical methods}


\maketitle
\flushbottom


\section{Introduction}\label{sec:introduction}

Streak cameras are widely used instruments in the particle accelerator community for the acquisition of time-resolved signals \cite{AWAKE:2022kmf, Bachmann:2020eoj, Hogan_2003, Rieger_2017, Bachmann_2020, Rossa_1992, Suntao_2018, Oz_2016}. The operational principle involves incident light signals, commonly photons, impinging on a cathode, thereby generating photoelectrons. These photoelectrons are then accelerated and then deflected vertically by an angle which depends on their arrival time. Deflected photoelectrons emit a light signal upon reaching a phosphor screen. This signal, captured by a CMOS camera, enables timing measurements of incident signals based on their vertical image position. The typical image reconstruction process is based on the light signal's amplitude as measured by the CMOS camera. In this article, we describe a novel algorithm for reconstructing the photoelectron distribution based on the recorded camera image. This is accomplished by initially identifying the characteristics of the signals generated by individual photoelectrons, and then decomposing the overall image in term of photoelectrons. This makes an analysis possible based on standard statistical techniques, so that uncertainty estimation on a picture-by-picture basis is available.

We have performed our studies in the context of the AWAKE experimental setup~\cite{Caldwell:2015rkk,AWAKE:2015taz,AWAKE:2022aeo,Batsch:2019vsw}. The AWAKE Collaboration pursues the demonstration of electron acceleration in a plasma wakefield driven by protons. In the context of AWAKE, streak cameras are used for capturing time-resolved images of the proton beam as well as the produced plasma wakefield structure. The resulting time-resolved images are used for studying a variety of features of the modulation of the proton bunch in the plasma~\cite{AWAKE:2018wrd,AWAKE:2020stp,AWAKE:2022kmf,Nechaeva_2023hosing,Batsch:2019vsw}. An important component of the setup is a high-power laser~\cite{Fedosseev:2016ccm}. This laser produces a high-intensity pulse that ionizes a 10-m-long column of Rb vapor, generating the plasma, and a timing reference signal that is sent to the streak camera. An additional specially prepared path for the laser is utilized for the studies described in this article.

In the following, we describe the experimental setup, with a dedicated focus on the streak camera. The specific setup of the laser system that was employed for the analyzed data is described in~\cite{Batsch:2019vsw} and briefly reviewed. We then describe the method developed for our study and show an example of image deconvolution. The timing reconstruction based on the deconvolved image is then described, and results are shown on the time reconstruction and uncertainty estimate.


\section{Experimental Setup}\label{sec:experimental_setup}

\subsection{Operational Principles of Hamamatsu Streak Camera C10910}

The Hamamatsu Streak Camera C10910 is the imaging instrument used to record images for the datasets of this paper~\cite{Hamamatsu:2022}. The streak camera is designed to detect photons ranging from the UV to the near-infrared spectrum and records them as 2D images with one spatial dimension and the temporal dimension as axes. We denote the temporal dimension as the vertical dimension on the resulting images.

\begin{figure*}[!ht]
  \includegraphics[width=\textwidth]{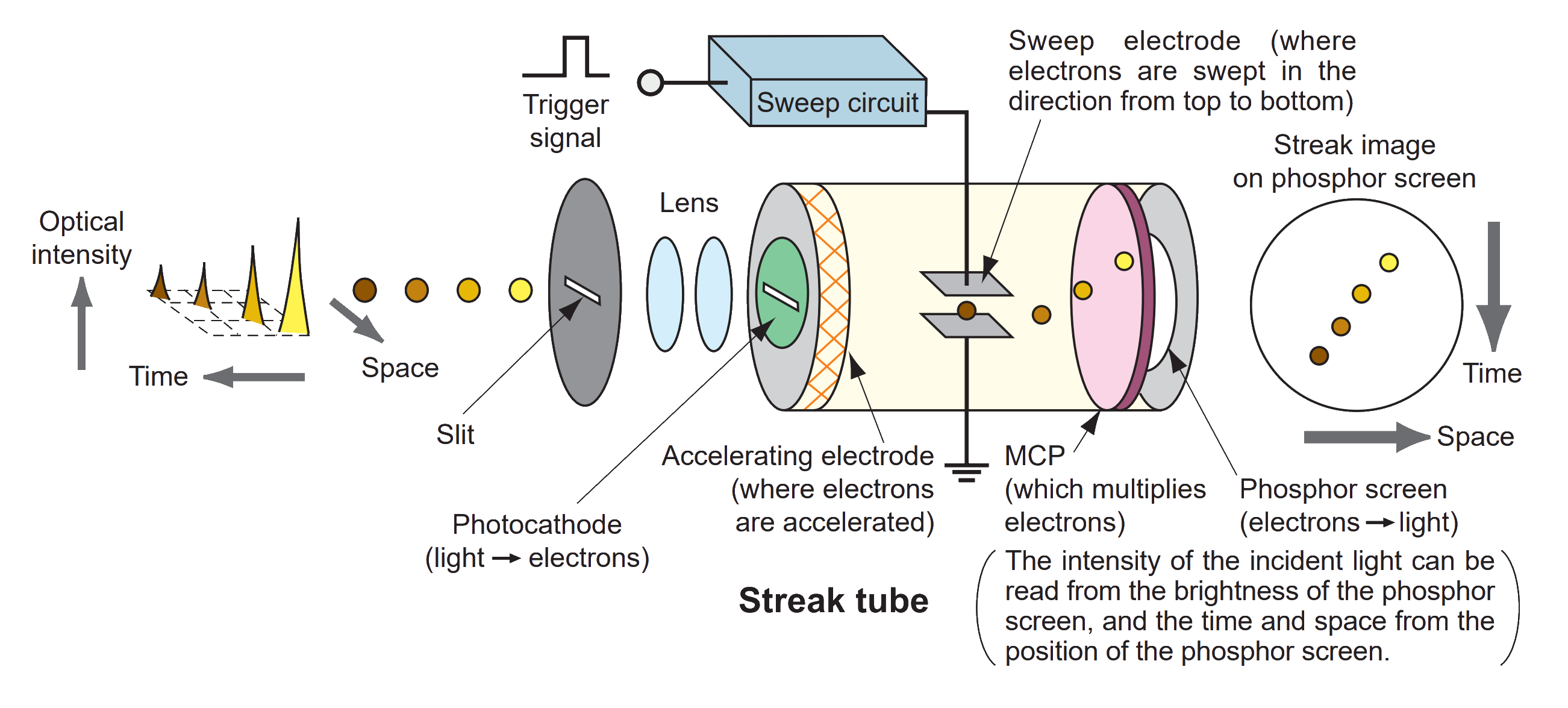}
  \caption{Illustration of the working principle of the Hamamatsu Streak Camera C10910. The diagram shows four light pulses with varying amplitudes and separated in time passing through the components of the streak camera before reaching the phosphor screen. This figure is taken from the Hamamatsu Manual \cite{Hamamatsu:2022}.}
  \label{fig:hamamatsu-streak-camera}
\end{figure*}

The operating principle of the streak camera, as illustrated in \autoref{fig:hamamatsu-streak-camera}, involves several stages. Initially, the light intersects with the a narrow slit whose width is set to 20~$\mu m$. Under the guidance of the imaging system, this light is directed to the photocathode, which converts a fraction of the incoming photons into photoelectrons. These photoelectrons are then accelerated towards the end of the streak tube and deflected in the direction perpendicular to the slit. After amplification, they impinge on the phosphor screen, generating light pulses. The deflection process's speed is adjustable and was set such that the time span of an image corresponds to either $73$~ps or $210$~ps for the analysed data.

The phosphorescent images are captured by a CMOS readout camera, which is located behind the phosphor screen. Due to the multiplication of the photoelectrons by the multi-channel plate (MCP) and the light production and emission process of the phosphor screen, individual photoelectons are observed as clusters spanning multiple pixels in the CMOS readout camera. The camera has 512 pixels in the vertical (time) direction and 672 pixels in the horizontal direction. One pixel in the vertical direction therefore corresponds to $0.143 (0.410)$~ps for the $73 (210)$~ps image time windows.

The streak camera is triggered by an external signal, and this is used to initiate the voltage sweep across the sweep electrode. In previous studies, the time jitter of the triggering mechanism has been determined to be 4.8~ps~(rms) \cite{Batsch:2019vsw}, implying that the resulting images jitter in the vertical dimension (time dimension) on a picture-by-picture basis by at least this amount. The intrinsic time resolution of the streak camera is much better than 4.8~ps so that relative time measurements amongst components of a single image can be determined with higher precision. Our algorithm aims to extract relative time measurements with the highest achievable precision using the streak camera.

\subsection{Laser System}

\begin{figure*}[!ht]
  \centering
  \includegraphics[width=0.6\textwidth]{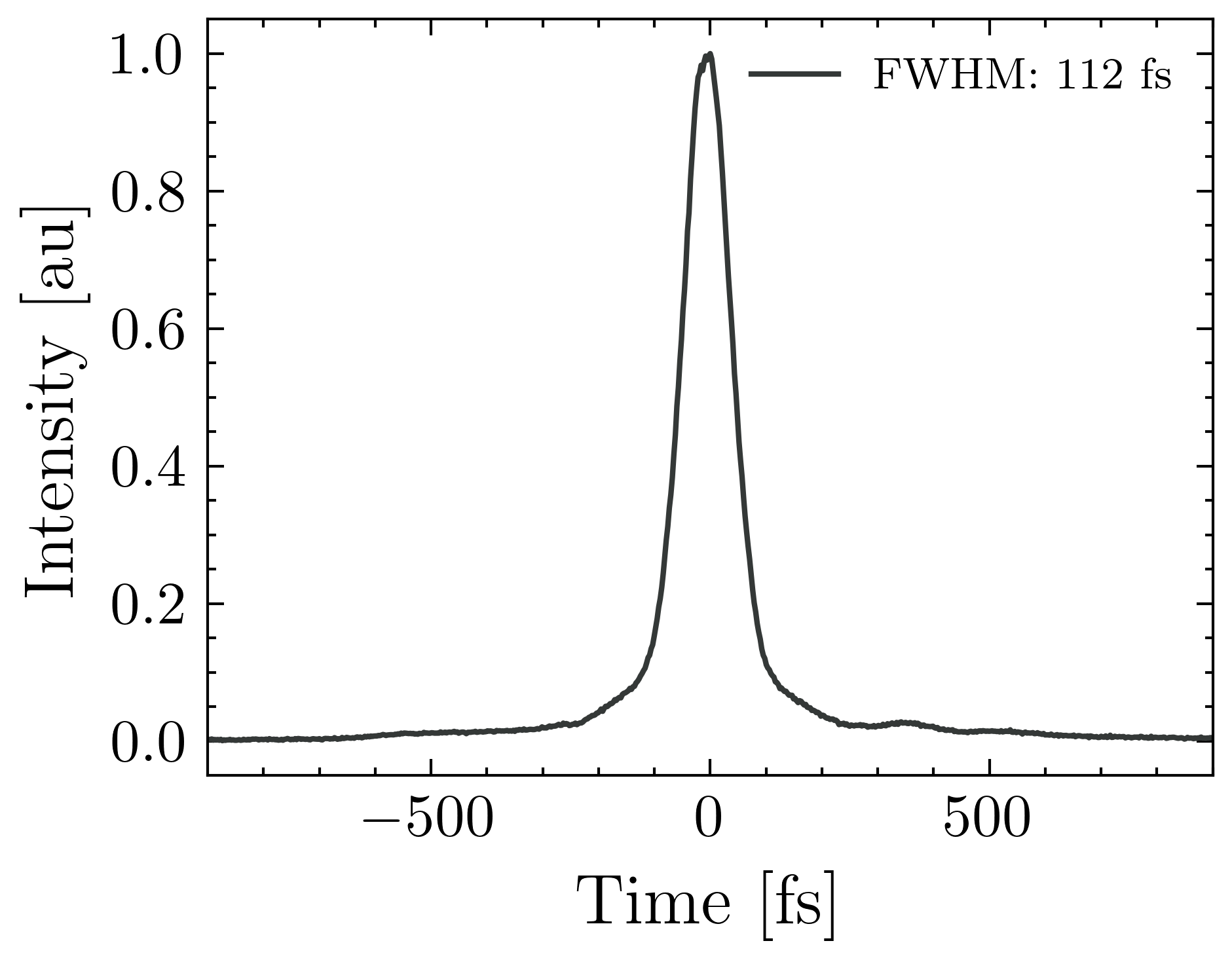}
  \caption{Temporal profile of the laser pulse obtained by a second-harmonic correlator}
  \label{fig:laser-pulse}
\end{figure*}

The data used for the development of our algorithm was generated by a TW-class Ti:Sapphire laser system~\cite{Batsch:2019vsw}. The laser system amplifies an erbium-doped fiber oscillator with frequency-doubled 780 nm output to produce pulses with an energy up to 650 mJ at 10Hz and a duration of 110 fs. 
For the operation of these studies, an attenuated  compressed laser pulse was used. The laser pulse duration shown \autoref{fig:laser-pulse} was obtained by a second-harmonic auto-correlator (Bonsai from Amplitude Company). More information on the laser system can be found in~\cite{Wing:2643452}.

\subsection{Light Reflection Setup}

The primary dataset used in the development of our streak camera signal reconstruction algorithm was acquired in the test setup described in~\cite{Batsch:2019vsw}. The experimental setup was a subsystem of the AWAKE experiment.

\begin{figure}[!ht]
  \centering
  {\includegraphics[width=0.63\textwidth]{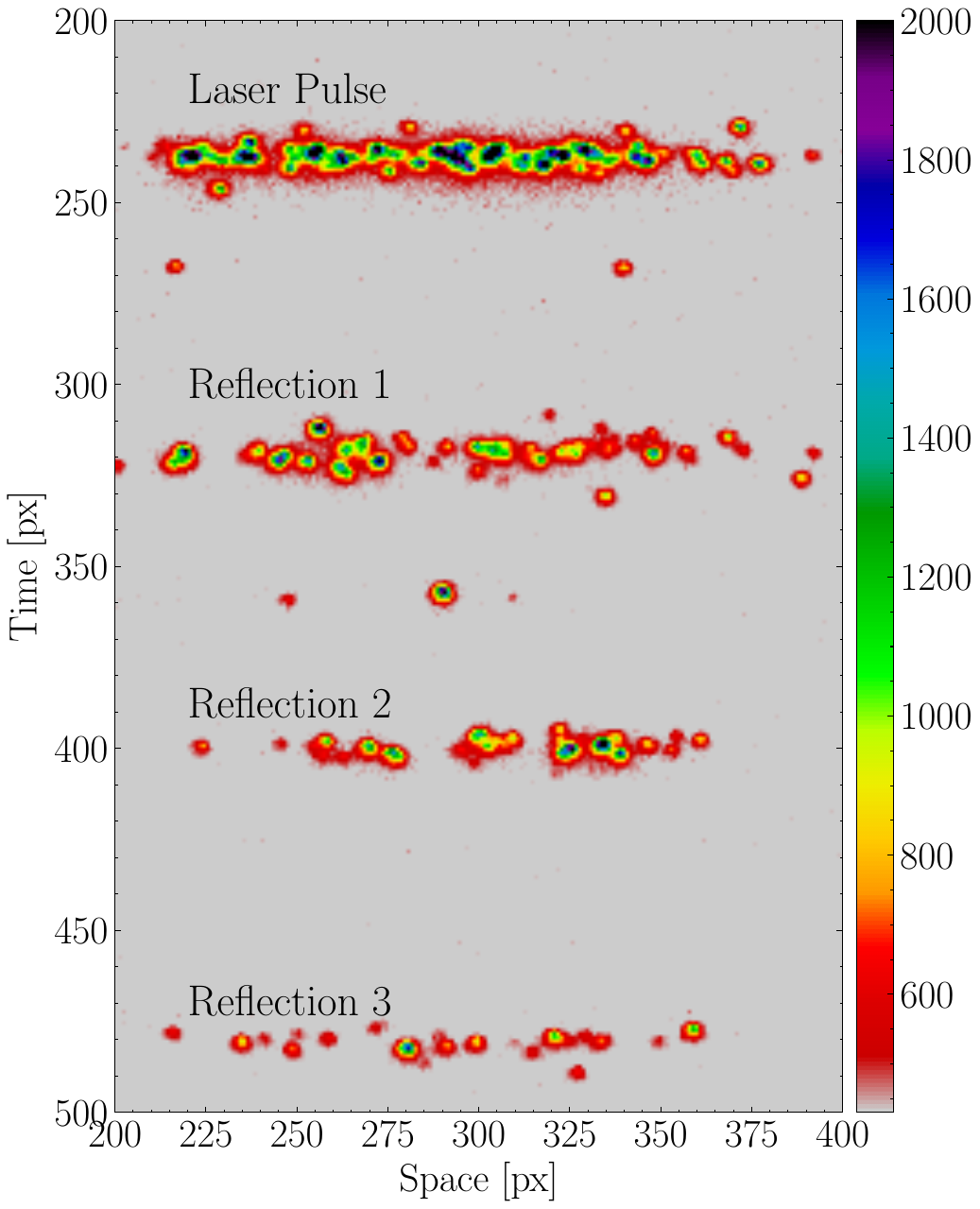}}
  \caption{Streak camera image of the primary laser pulse and the first three reflections inside the glass plate.}
  \label{fig:streak-image}
\end{figure}

The laser light path to the streak camera included a~$\sim 5$~mm-thick piece of glass to obtain multiple surface reflections with fixed time intervals. In total, 1070 laser reflection images were acquired with this setup. The sweep time used for this dataset\footnote{We note that the sweep time reported in~\cite{Batsch:2019vsw} is not correct.} was $210$~ps, so that images resulting from internal reflections in the glass are expected to be separated by $\sim 40$~ps, equivalent to $\sim 100$~pixels~(px). \autoref{fig:streak-image} shows a streak camera image from this data set. We select a 200-px-wide region of interest in the spatial direction to focus on the laser pulse and its reflections.Within the ROI, we observe the primary laser pulse at approximately pixel coordinate $237$ on the time axis, along with consecutive reflection pulses generated from up to three round-trips through the glass at around time pixel coordinates of $318$,~$400$,~and $480$, respectively. We convert from pixel coordinate to time in picoseconds as described below.

The expected delay between the reflections could not be determined precisely based on measurements of the glass thickness and its refraction index. However, by comparing the measured time differences between different reflections, we are able to determine the uncertainty of the time measurements achievable by the streak camera, without knowing the exact time spacing between the reflections.

When analysing the obtained streak camera images, we observed that the isolated signals in the streak camera images typically appeared as either single pixels or as small pixel clusters with pixel amplitudes near the CMOS baseline. The single pixels were interpreted as noise fluctuations, whereas the small clusters were identified as clusters generated by single photoelectrons. The features of the clusters are presumably affected by numerous factors such as the MCP pore size, the amplification settings of the MCP, the image spread created in the phosphor screen, the CMOS sensors pixel size, etc. These factors affected the achievable time resolution of the streak camera. In order to apply the algorithm described below to data sets with different streak camera operational settings, the photoelectron characterization steps will need to be repeated.

In our study of images taken with different sweep speeds, we observed that the streak camera time-window size has little effect on the pixel size of clusters produced by single photoelectrons. As a consequence, the algorithm operates at the pixel level and as the cluster size in pixels remains unchanged across different sweep speeds.

\subsection{Additional Dataset}

We have also made use of an additional dataset with $73$~ps sweep speed. This allowed us to compare images acquired with different sweep speeds, and provided separate calibration data for our algorithms, as explained below. The laser pulse, used as a timing reference in this dataset, did not include the glass piece and contained no reflections of the laser pulse. An artificial $0$-distance reflection was simulated in the data by dividing the recorded laser pulse image into two sets of pixels along the spatial direction. In this case, we expect the same time to be measured in each half and a comparison of the recorded times allows for a study of the uncertainties in the time reconstruction. A total of 200 images were recorded in this set of conditions.


\section{Photoelectron Decomposition Method}\label{sec:photoelectron-decomposition-method}

\begin{figure*}[!ht]
  \includegraphics[width=\textwidth]{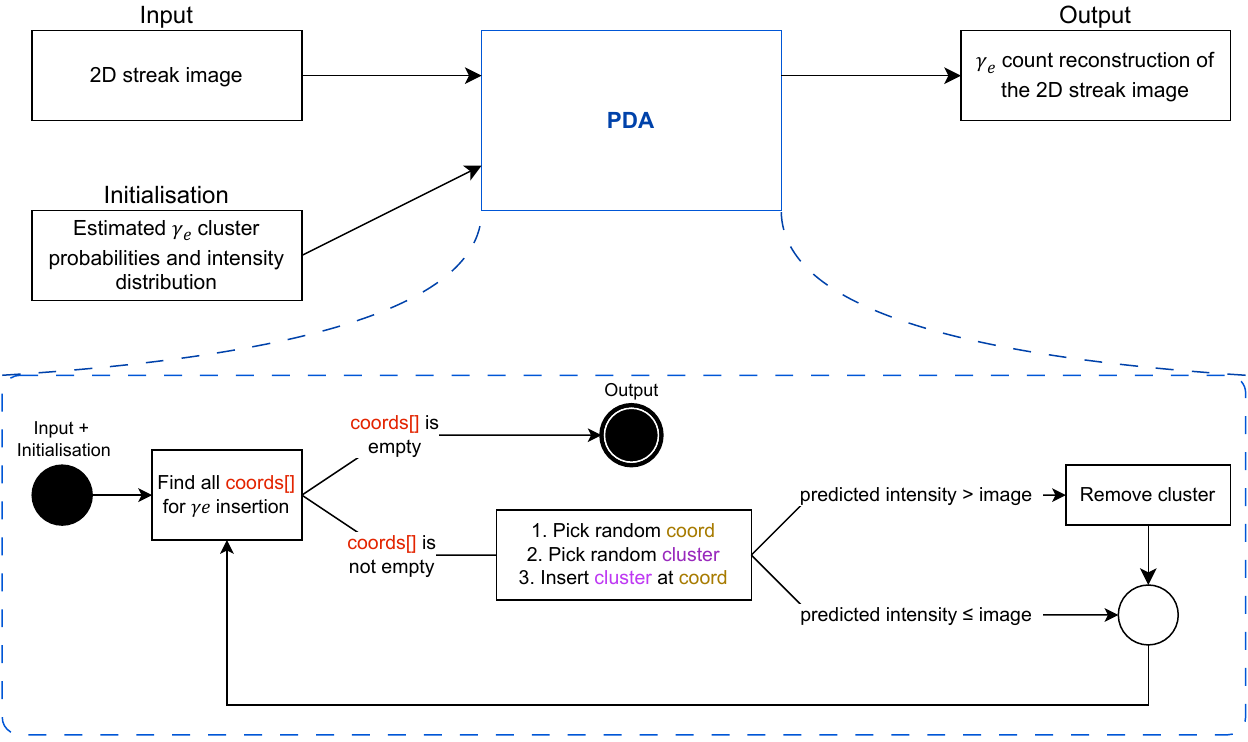}
  \caption[]{Overview of the Photoelectron Decomposition Algorithm. The algorithm takes as input the recorded image and the photoelectron cluster probabilities. The algorithm then performs a reconstruction of the photoelectron distribution for the recorded image. The symbol $\gamma_e$ stands for photoelectron.}
  \label{fig:algorithm}
\end{figure*}

An overview of our Photoelectron Decomposition Algorithm (PDA) is given in \autoref{fig:algorithm}.  In the following subsections, we describe the different steps in the algorithm.  The performance is then described in the following section.

\FloatBarrier

\FloatBarrier
\subsection{Isolating Individual Photoelectrons Using Binary Filtering}

\begin{figure}[!ht]
  \centering
  {\includegraphics[width=\textwidth]{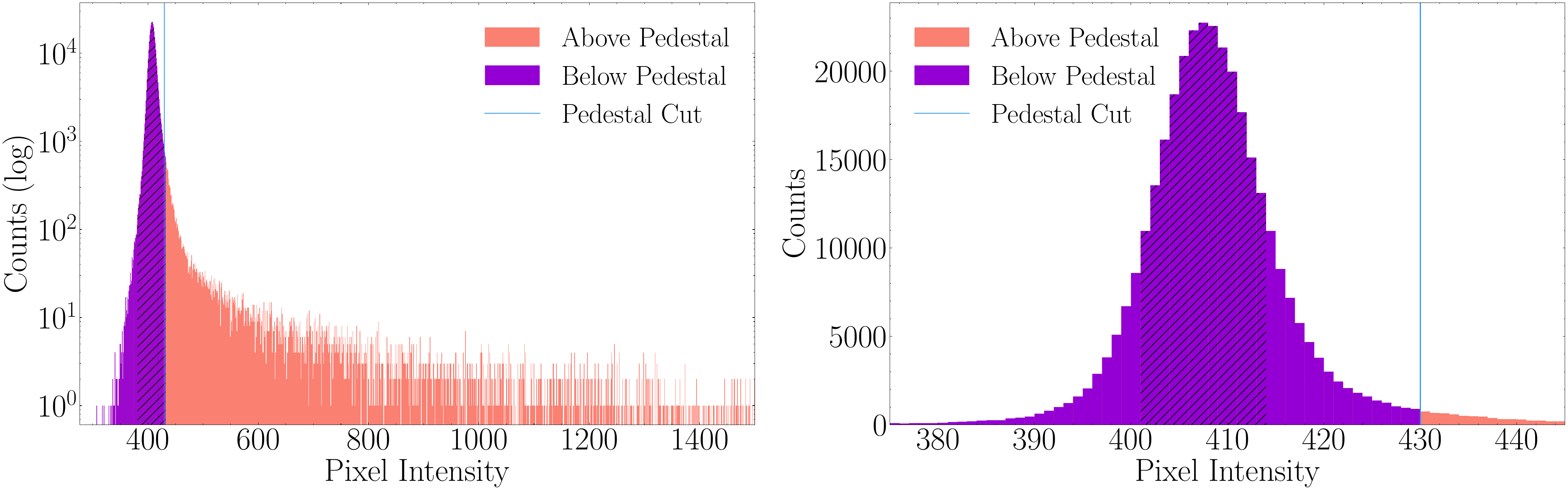}}
  \caption{Histogram of the pixel amplitudes for one image in our dataset. The left panel shows the full distribution in log scale, while the right panel shows a blow-up of the pedestal region using linear scale from a regions outside the ROI. The vertical line indicates the threshold value of the pedestal used for binary filtering.}
  \label{fig:background-histogram}
\end{figure}

In a first step, we employ a binary filter to isolate as many single-photoelecton induced clusters as possible from the streak camera image. To achieve this, we apply a cutoff determined from the pedestal value of the streak camera image. The individual pixel amplitude distribution for one image in our dataset is shown in~\autoref{fig:background-histogram}. A blow-up of the  distribution around the pedestal value taken from a streak camera image region outside the ROI is also shown. We chose a threshold amplitude value of 430 counts for the binary filtering. This allows a reduction of the bulk of the pedestal fluctuations while preserving the bulk of photoelectron generated signals.

\begin{figure*}[!ht]
  \centering
  \includegraphics[width=0.55\textwidth]{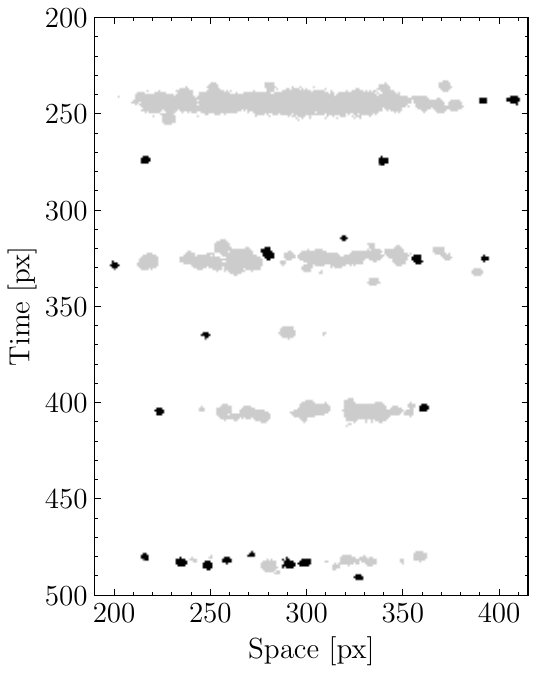}
  \caption[]{Pixels in the image shown in \autoref{fig:streak-image} passing the binary thresholding test are shown. The pixels belonging to clusters identified as resulting from single photoelectron are shown in the darker shading, while the lighter shading represents pixels from clusters that are too big or too small to be identified as resulting from single photoelectrons.}
  \label{fig:filtered-binary-masks}
  \vspace{-15px}
\end{figure*}

The operation of the binary filter~\cite{Gonzalez_2008} is rather straightforward. It assigns pixel values of 1(0), depending on whether their amplitude is above(below) a predefined threshold. We demonstrate this filtering on the streak camera image shown in \autoref{fig:streak-image} as a filtered image in \autoref{fig:filtered-binary-masks}. All pixels above the threshold are displayed against a white background. The two shadings (black and grey) are used to distinguish clusters associated with single photoelectrons as explained below.

\begin{figure}[!t]
  \centering
  {\includegraphics[width=0.8\textwidth]{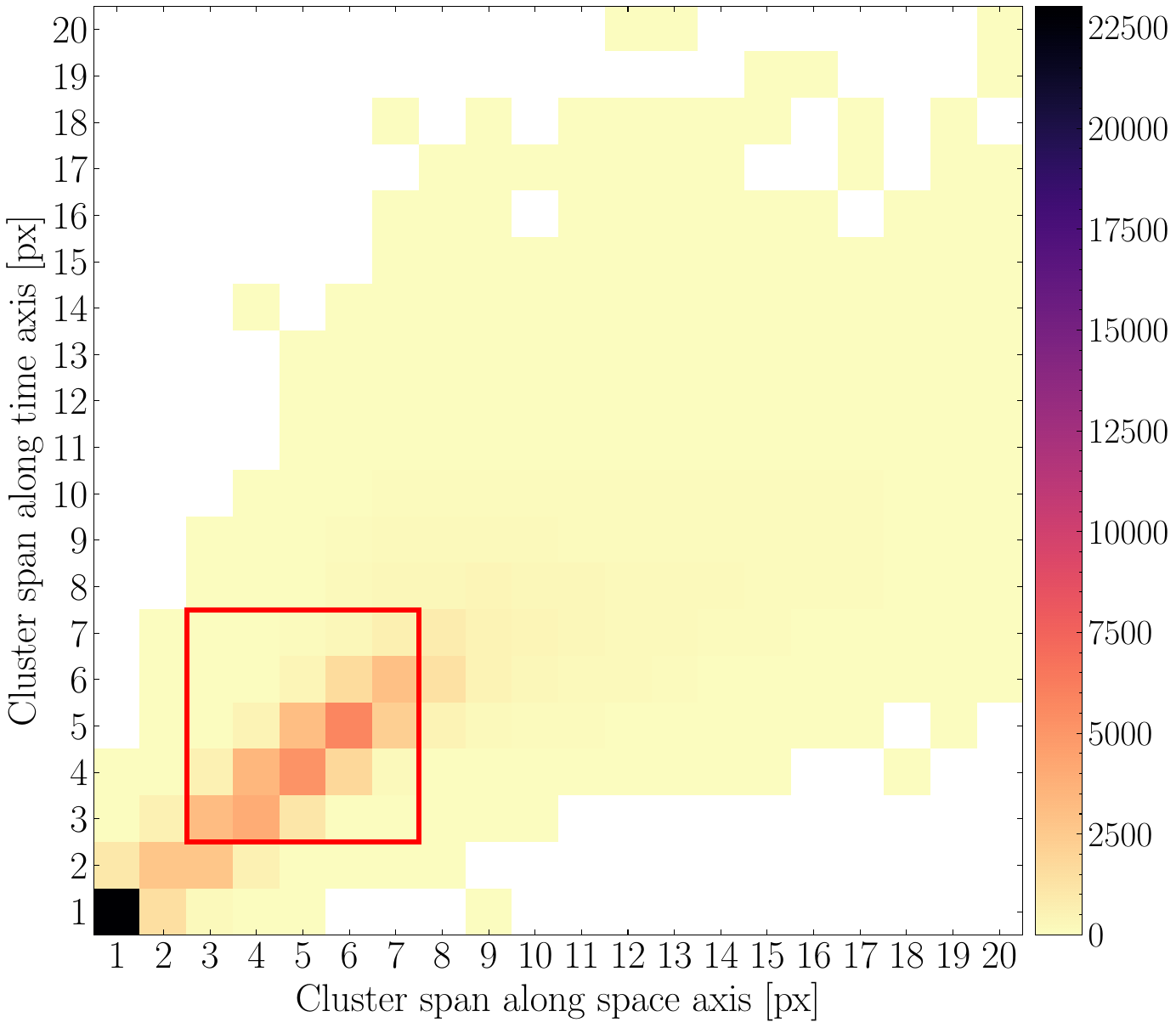}}
  \caption{Heatmap of the cluster sizes along space and time axes of the streak image. The color scale gives the number of clusters found with a particular size. The red box indicates the cluster size region used to identify single photoelectrons.}
  \label{fig:cluster-heatmap}
\end{figure}

The next step is to apply a connected component labeling algorithm~\cite{Rosenfeld:1966} to the pixels with value 1. This algorithm identifies and groups all pixels with neighboring non-zero pixels along the temporal and spatial axes into clusters and assigns a unique label to each cluster, effectively forming distinct, traceable entities.
The resulting cluster size distribution along the spatial and temporal axes are shown in \autoref{fig:cluster-heatmap}. Two clear and identifiable  cluster size regions are visible.  The peak at a cluster size of one pixel in the temporal and spatial directions is from noisy pixels with amplitude above threshold.  We identify the larger island with $\sim 5$ pixels along the spatial and temporal axes as arising from single photoelectrons.

To select clusters identified as arising from single photoelectrons, we employ multiple limitations on the cluster size. Firstly, we restricted the cluster size to be between 3 to 7 pixels in both the spatial and temporal axes as shown with the red square in \autoref{fig:cluster-heatmap}. Secondly, we discarded any cluster that has less than 8 total pixels above the pedestal. This requirement was needed to eliminate a special class of noise clusters appearing as a diagonal feature in the image. Lastly, we discarded the cluster if any pixel had an amplitude more than 500 counts above the pedestal threshold, again to eliminate clusters deemed to be caused by noise.

\autoref{fig:filtered-binary-masks} shows the result of this filtering on a sample image, where the black pixels are the pixels that passed the filtering and the grey pixels are the pixels that failed the filtering. From the 1070 images in our primary dataset, we detected 23888 single photoelectron clusters. This significant dataset allowed for a detailed study clusters identified as resulting from single photoelectrons.

\FloatBarrier

\clearpage
\subsection{Analysis of Individual Photoelectron Clusters}

To analyze the amplitude distribution caused by individual photoelectrons, all clusters from the laser reflection images were examined. The two-dimensional clusters were transformed into one-dimensional clusters by summing pixels along the space axis. We then categorized the clusters based on their width along the time axis. As a result of this process we obtained five separate classes each with different pixel widths. Additionally, for each category we assigned class probabilities by dividing the number of observed clusters in the class by the total number of observed clusters. \autoref{tab:cluster-size-distribution}
shows associated counts and probabilities for each category. $95.3$~\% of clusters identified as resulting from single photoelectrons have widths along the time axis of $3-5$ pixels.

\begin{table}[h]
  \centering
  \begin{tabular}{|c|c|c|c|c|c|}
    \hline
    \textbf{\textit{Cluster Class}} & \textbf{3 px} & \textbf{4 px} & \textbf{5 px} & \textbf{6 px} & \textbf{7 px} \\
    \hline
    Detected Count                  & 6099          & 11497         & 6283          & 881           & 309           \\
    \hline
    Probability                     & 0.243         & 0.459         & 0.251         & 0.035         & 0.012         \\
    \hline
  \end{tabular}
  \caption{Photoelectron cluster size distribution along the time axis. `Detected Count' gives the number of clusters detected for each cluster size over 1070 images and `Probability' gives the associated probabilities for each cluster size.}
  \label{tab:cluster-size-distribution}
\end{table}

\begin{figure*}[!ht]
  \centering
  \includegraphics[width=\textwidth]{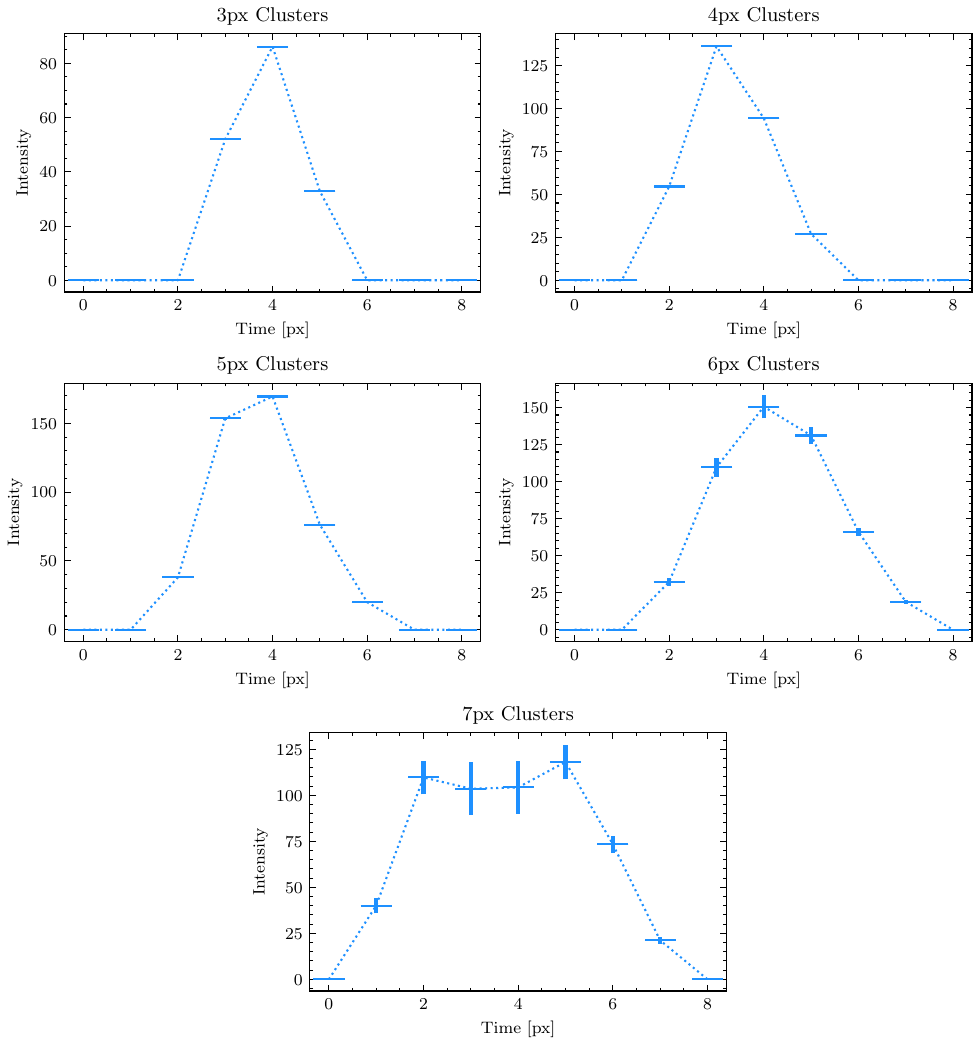}
  \caption{Averaged time profiles of clusters identified as resulting from single photoelectrons, separated into clusters with different numbers of pixels as indicated in the legends The amplitude values are averaged and centered within 9-pixel windows. The error bars represent the RMS of the amplitudes recorded at the different cluster pixel locations divided by the square root of the number of clusters in that class.}
  \label{fig:average_amplitude_profile}
  \vspace{-15px}
\end{figure*}

For each of these classes, we calculated an average amplitude profile along the time axis and its fluctuations. These profiles are shown in \autoref{fig:average_amplitude_profile} for each class. The profiles were centered within 9-pixel windows to allow for a visual comparison between the different classes. The error bar is the uncertainty on the average calculated from the RMS divided by the square root of the number of clusters in that class. We next use these cluster profiles and their occurence probabilities to decompose streak camera images into distributions of photoelectrons.

\FloatBarrier
\clearpage
\subsection{Photoelectron decomposition algorithm (PDA)}

\begin{algorithm}[!ht]
  \hrule
  \vskip 0.5cm

  \SetAlgoLined
  \SetKwInOut{Input}{Input }
  \SetKwInOut{Output}{Output }

  \Input{Amplitude distribution of clusters along the time axis}

  \Output{Individual photoelectron counts}

  \BlankLine
  \SetKwFunction{FMain}{reconstructPhotoelectrons}
  \SetKwProg{Fn}{Function}{:}{}

  \Fn{\FMain{amplitude\_distribution}}{
    cluster\_probabilities[] $\leftarrow$ Probabilities for photoelecton clusters of different widths\;
    predicted\_amplitude[] $\leftarrow$ Zeros with size equal to amplitude\_distribution\;
    photoelectron\_distribution[] $\leftarrow$ Zeros with size equal to amplitude\_distribution\;
    insertion\_locations[] $\leftarrow$ find\_photoelecton\_insertion\_locations(predicted\_amplitude[])\;
    \BlankLine
    \While{sum(insertion\_locations) $\neq$ 0}{
      location $\leftarrow$ pick\_random\_location(insertion\_locations[])\;
      cluster[] $\leftarrow$ pick\_random\_cluster(cluster\_probabilities[])\;
      next\_amplitude\_prediction[] $\leftarrow$ insert\_cluster(predicted\_amplitude[], location, cluster[])\;

      \If{all(next\_amplitude\_prediction[] $\leq$ amplitude\_distribution[])}{
        photoelectron\_distribution[] $\leftarrow$ increment\_photoelectron(location)\;
        predicted\_amplitude[] $\leftarrow$ next\_amplitude\_prediction[]\;
      }
      insertion\_locations[] $\leftarrow$ find\_photoelecton\_insertion\_locations(predicted\_amplitude[])\;
    }
    \Return photoelectron\_distribution\;
  }
  \vskip 0.5cm

  \hrule
  \vskip 0.5cm

  \caption{Photoelectron Decomposition Algorithm}
  \label{alg:photoelectron-decomposition-algorithm}

\end{algorithm}

The photoelectron decomposition algorithm is applied to clusters that are larger in size than single photoelectrons.  These clusters are first summed along the spatial direction resulting in a one-dimensional amplitude distribution along the temporal axis. Our algorithm then builds up a simulated amplitude distribution based on summing amplitudes distributions from single photoelectrons placed at different locations along the temporal direction.  The algorithm is designed to have the simulated distribution reproduce the observed distribution. The end result of the process is a distribution of photoelectrons along the temporal direction.

The algorithm begins with an empty simulated amplitude distribution. A pixel coordinate for insertion of a photoelectron is chosen at random, and the cluster size is then randomly chosen according to the probabilities given in \autoref{tab:cluster-size-distribution}. An attempt is then made to add the average amplitude profile to the simulated distribution.
In this process, the photoelectron amplitude profile is centered as follows: if the cluster has an odd number of pixels, the amplitude profile is centered on the chosen pixel location, while, if the cluster has an even number of pixels, the profile is randomly centered on one of the two central pixels with equal probability for each. If the insertion results in amplitudes greater than the observed amplitude for any pixel, the insertion is rejected. Otherwise, the insertion is accepted and the algorithm moves to the next randomly picked location. The algorithm terminates when it cannot detect any possible location for photoelectron insertion. A pseudo-code of the algorithm is given in \autoref{alg:photoelectron-decomposition-algorithm}.

\FloatBarrier

\subsection{Reversible decomposition}

\begin{figure*}[!h]
  \centering
  \includegraphics[width=0.9\textwidth]{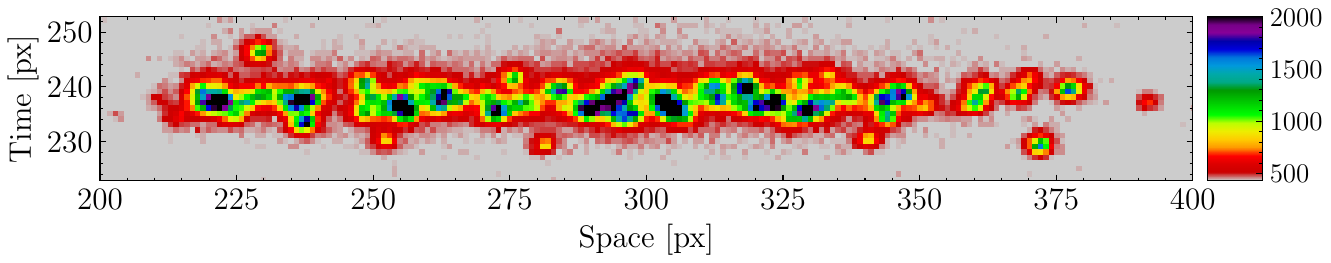}
  \caption{Expanded view of the signal region produced by a laser pulse. The signal amplitude is given in color scale. }
  \vspace{6pt}
  \label{fig:laser-roi}
\end{figure*}

It is possible to convert the pixel amplitudes to photoelectron counts and vice versa. This provides a valuable check for our algorithm. As an example, we show a detailed view in the 30-pixel-wide (along the time axis) window containing a laser pulse image in \autoref{fig:laser-roi} and proceed to analyze this image. As described above, our algorithm works in a stochastic manner, and successive decompositions will produce different photoelectron distributions.

\begin{figure*}[!ht]
  \centering
  \includegraphics[width=\textwidth]{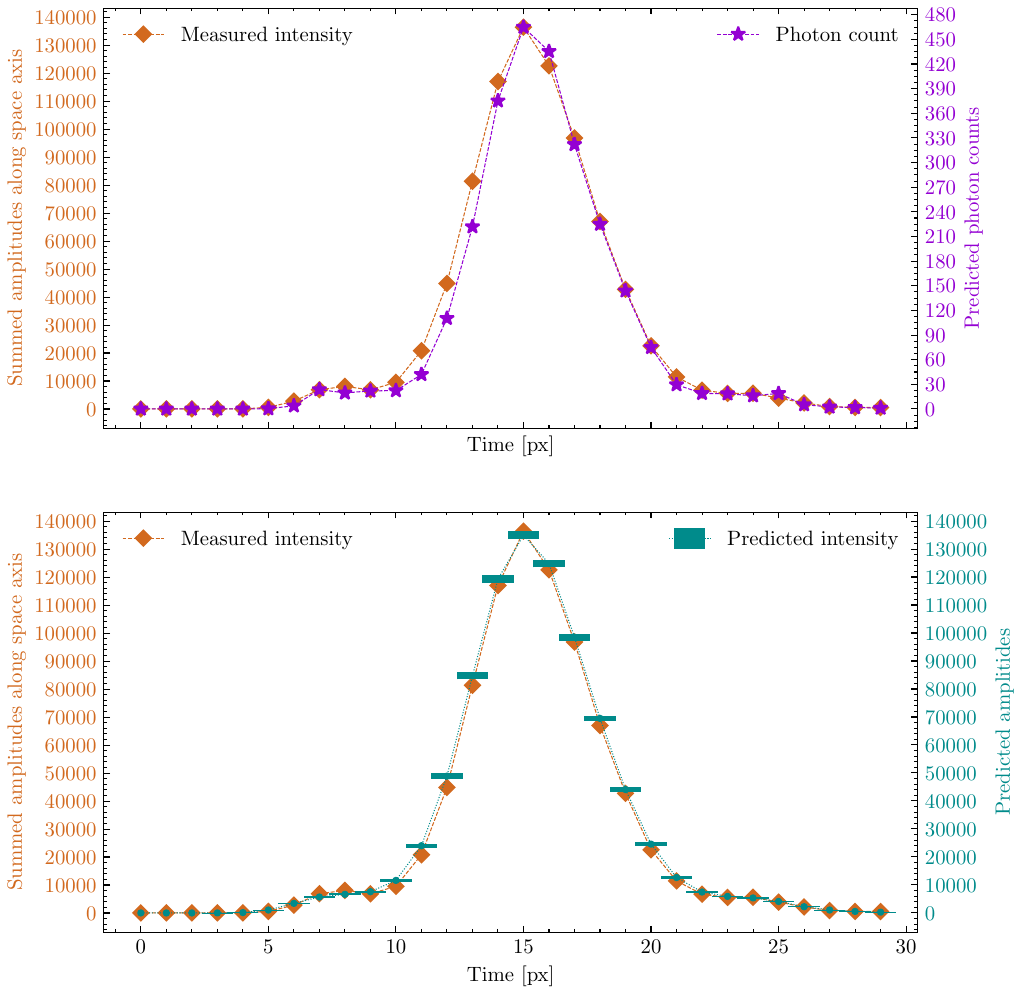}
  \caption{
    Upper panel: Measured time profile and extracted number of photoelectrons for the image shown in \autoref{fig:laser-roi}. The horizontal axis shows time in pixel units, the left hand side vertical axis gives the measured amplitudes and the right hand side vertical axis gives the extracted photoelectron counts. Lower panel: Comparison between predicted amplitudes to the measured amplitudes using our reversed photoelectron decomposition algorithm. The error bars show the RMS of the predicted amplitudes.}
  \label{fig:laser-roi-sum-along-space-vs-predicted-photons}
\end{figure*}

The amplitude distribution for this image, integrated over the spatial direction, is shown in the upper panel in \autoref{fig:laser-roi-sum-along-space-vs-predicted-photons}. The result of the photoelectron decomposition algorithm is also shown. As seen in the Figure, the photoelectron distribution is somewhat narrower than the observed amplitude distribution, as the photoelectron clusters broaden the image. The decomposition process effectively deconvoles the amplitude distribution and produces a somewhat narrower distribution.

We can perform the procedure in reverse by using the photoelectron cluster distribution, selecting from our cluster spectra according to their probabilities, and creating an amplitude distribution. We have repeated the process of recreating the amplitude distribution 10000 times for the image shown in the upper panel of \autoref{fig:laser-roi-sum-along-space-vs-predicted-photons}, based on one photoelectron decomposition. The results are shown in the lower panel of the figure, where the observed amplitude distribution is given by the diamond shaped symbols and the average reconstructed image is shown in horizontal bars. The RMS of the amplitude distributions from the 10000 repetitions is shown as the error bar on the prediction and is typically within the size of the symbol. As can be seen, the average agrees well with the observed distribution, and the spread in the predicted signal shape is quite small.


\section{Timing measurements}\label{sec:timing-measurements}
The primary use of the streak camera is to extract timing information. We have studied the potential timing resolution of features within individual streak images. This has been accomplished through the usage of the photoelectron decomposition method on two distinct datasets: the `reflections' dataset, which used a $210$~ps sweep time, and a dataset with $73$~ps sweep time.

\subsection{Finding the location of features in time}

Our algorithm provides the pixel location of photoelectrons along the time axis of the image and thereby enables the application of standard statistical methods to estimate uncertainties for timing feature extraction. Typical approaches for finding feature location, e.g., fitting a Gaussian function to the amplitude profile of the feature, does not allow the calculation of the uncertainty on the time measurement on a per image basis or a per feature basis.

We calculate a feature time using a weighted average of the photoelectron pixel locations, where the weights are the photoelectron counts. The weighted average is given by

\begin{align}
  \overline{t}_{px} = \frac{\sum_{i=1}^{n} w_i t_i}{\sum_{i=1}^{n} w_i},
\end{align}\label{equation:weighted-average}

\noindent where $t_i$ is the pixel number of the $i$-th pixel along the time axis in the image and $w_i$ is corresponding photoelectron count. The averaged pixel value is then converted to time using the conversion based on the sweep time window. The uncertainty on this time estimate is expected to scale inversely with the square root of the total number of photoelectrons.

We use this scheme to analyze the data from our two datasets. As a first step, we need to locate a ROI within the image to be analyzed. We do this by fitting Gaussian functions to temporal amplitude profiles summed among the space axis and choosing a suitably broad region around the peak pixel. The ROI shown in \autoref{fig:laser-roi} was chosen by adding $\pm 15$~px to the center pixel from the Gaussian fit. We have repeated this process for the ROIs found in each image of our datasets and applied our time reconstruction algorithm to these ROIs.

\begin{figure}[!ht]
  \centering
  {\includegraphics[width=\textwidth]{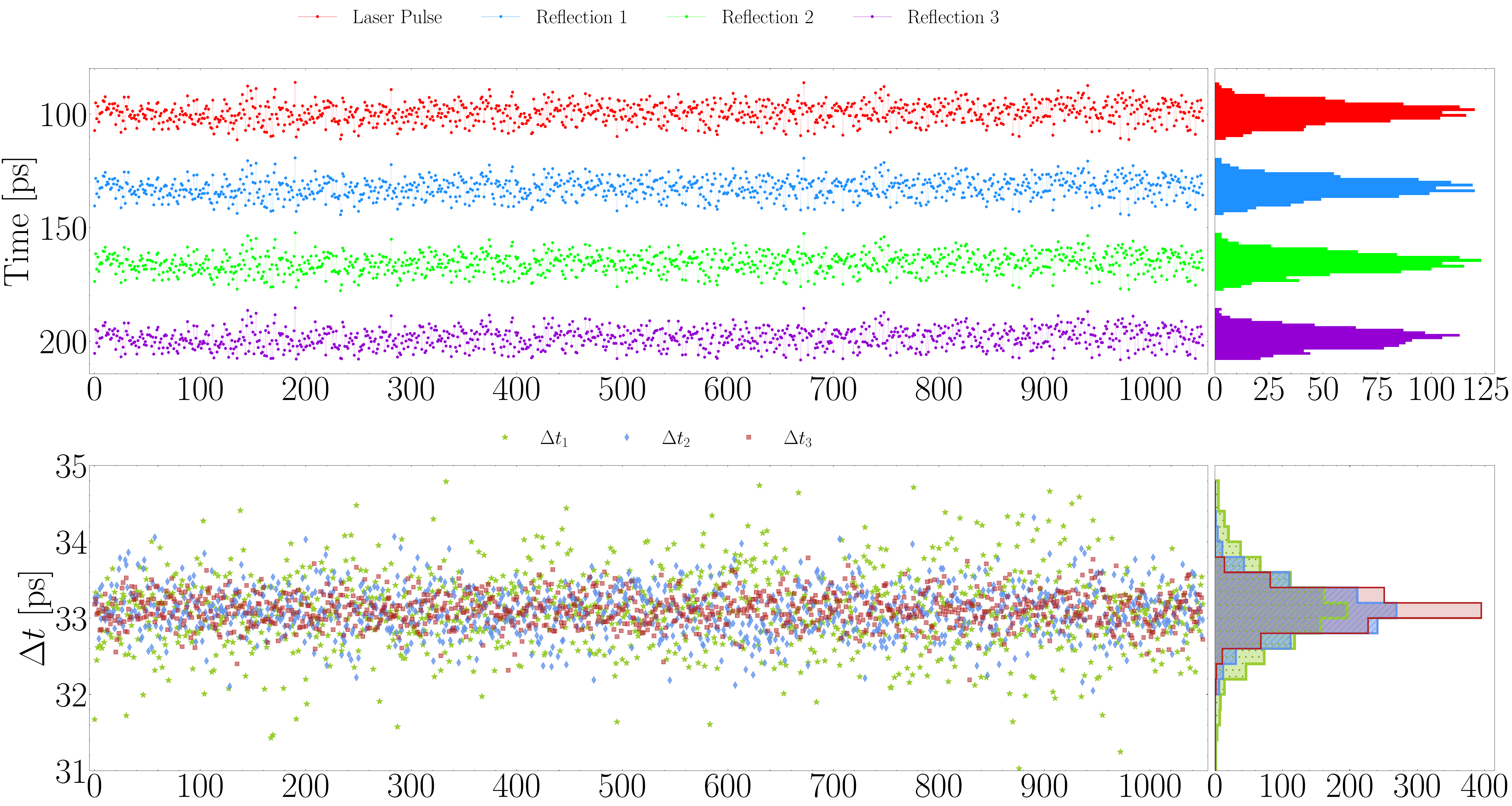}}
  \caption{Top panel: Reconstructed timings for the laser pulse and its three reflections (1-3) from multiple round trips through the glass slab, displaying all images from the reflection dataset. Bottom panel: Time differences between the measured arrival times of the laser pulse and its subsequent three reflections.}
  \label{fig:all_events_mlp_time}
\end{figure}

In the upper panel of \autoref{fig:all_events_mlp_time} we show the reconstructed times as a function of image number for all 1070 images in our glass pane reflection dataset. The top row of times are for the laser pulse, while the three rows below are from (1-3) sets of consequtive reflections in the glass. The variation of the recorded times of each pulse are due the trigger jitter of the streak camera. This is shown by the fact that the fluctuations of each pulse are correlated with all of its reflections.

We use the notation $\Delta{t_1}$ to represent the time difference between the laser pulse and Reflection 1. Similarly, $\Delta{t_2}, \Delta{t_3}$ give the time differences between the reflections with 1 and 2 round trips, and the reflections with 2 and 3 round trips in the glass, respectively. The mean values are
\begin{eqnarray*}
  \overline{\Delta{t_1}} &=& 80.72 px (33.11 ps) \;\;\; RMS= 0.53 px (0.22 ps) \\
  \overline{\Delta{t_2}} &=& 80.69 px (33.10 ps) \;\;\; RMS= 0.77 px (0.31 ps) \\
  \overline{\Delta{t_3}} &=& 80.67 px (33.09 ps) \;\;\; RMS= 1.31 px (0.54 ps) \;,
\end{eqnarray*}\label{eq:time-differences}
in excellent agreement with each other.

We observed that the time differences between the laser pulse and its reflections are measured with much higher precision than the absolute times of the features with respect to the streak camera trigger, which is in agreement with the $\sim 4.8$~ps~(rms) jitter found in \cite{Batsch:2019vsw}.

\subsection{Time resolution}

We used the following simple parametrization for evaluating the uncertainty on the time measured for a given image feature:

\begin{equation}
  \label{Eq:resolution}
  \sigma_{t} = \frac{ \sigma_{\rm eff}}{ \sqrt{n_{\gamma}}}
\end{equation}

\noindent where $\sigma_{\rm eff}$ is an effective resolution per photoelectron and $n_{\gamma}=\sum_{i=1}^{n} w_i$ is the number of photoelectrons found in the feature for which we are extracting a time measurement. The parametrization presented in this work is applicable when $n_{\gamma}$ is neither too small nor too large. We have tested its effectiveness for $400 \lesssim n_{\gamma}\lesssim 3000$. In our experimental setup, the average number of photoelectrons reconstructed for the laser pulse was approximately 2800, which decreased by a factor of approximately 0.55 for each round-trip through the glass. The value for $\sigma_{\rm eff}$ was determined from the primary laser pulse by splitting the image into two parts along the space axis and
analyzing the resulting reconstructed time distribution. This procedure yielded $\sigma_{\rm eff}=18.28$~px. Although this is an effective per photoelectron resolution, it only applies when the number of photoelectrons is large enough as the smallest number of $n_{\gamma}$ used in our tests was approximately $400$.

\begin{figure*}[!ht]
  \centering
  \includegraphics[width=0.9\textwidth]{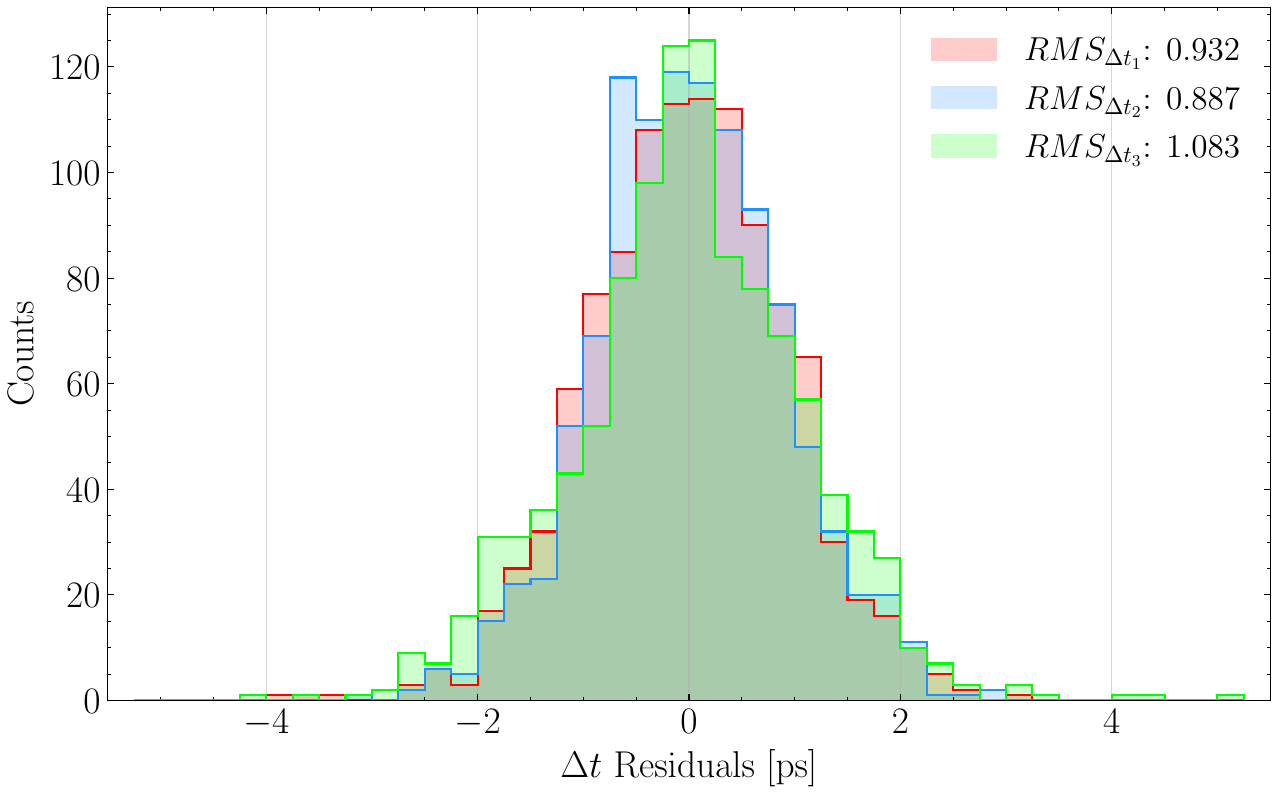}
  \caption[]{Residual distribution using the uncertainty parametrization \autoref{Eq:resolution}.}
  \label{fig:parametrisation-2-residuals}
\end{figure*}

As a test of our parametrization for the resolution, we calculated the residual distributions for the different $\Delta{t}$. The residual for $\Delta{t_1}$ for image $i$ is defined as
\begin{equation}
  r_{1,i} = \frac{\Delta{t_{1,i}}-\overline{\Delta{t_{1}}}}
  {\sqrt{\sigma_{t,i, \rm{Laser Pulse}}^2+\sigma_{t,i,\rm{Reflection 1}}^2}}
\end{equation}
where the index $i$ gives the image number.
$\overline{\Delta{t_{1}}}$ is the mean of the $\Delta{t_{1,i}}$ and was reported above. The corresponding residuals were calculated for $\Delta{t_2}$ and $\Delta{t_3}$. \autoref{fig:parametrisation-2-residuals} shows the residual distributions of all three $\Delta{t}$.

As can be seen in the figure, the residual distribution is centered on $0$ by construction, and has a width close to $1$, indicating that the uncertainty on the time measurement is well estimated.
Given that these results were extracted using a dataset with a $210$~ps sweep time window, we can report the timing resolution for this sweep window as

$$
  \sigma_t = \frac{7.50}{\sqrt{n_{\gamma}}}~ps \; .
$$
for $400 \leq n_{\gamma}\leq 3000$. I.e., a timing resolution of $140$~fs is reached for $n_{\gamma}=3000$ and this sweep window.

As discussed above, the resolution of the streak camera is due primarily to the single photoelectron cluster size in pixels, so that we expect a scaling of the resolution approximately with the length of the sweep time window. Using a $73$~ps time window, our parametrization then gives a time resolution of

$$
  \sigma_t = \frac{2.61}{\sqrt{n_{\gamma}}}~ps \; .
$$

We have tested this prediction by analyzing an AWAKE proton bunch modulation dataset with the streak camera window sweep time set to $73$~ps. The full photoelectron cluster characterization was carried out for the streak camera settings used in this dataset.  The photoelectron decomposition was then applied to the reference laser pulse in the acquired images. We used the technique of splitting the laser pulse image spatially in two parts and measuring the time difference found between the two halves of the pulse. The dataset consisted of 200 images and the typical number of photoelectrons reconstructed from the laser pulse was 800, with each half having approximately 400 photoelectrons.  This yielded a predicted resolution for each slice of the laser pulse of $130$~fs. Given that the uncertainties on the slices are expected to be uncorrelated, the uncertainty on the time difference between the two slices is expected to be $\sqrt{2}$ times larger than the uncertainty on each slice, which is approximately $180$~fs. We measured the RMS of the time difference between the slices of the laser pulse as $200$~fs, which agrees well with the predicted value, so that we are confident our parametrization of the time resolution in terms of pixel quantities can be transformed to a time resolution using the sweep time window.


\section{Conclusions}\label{sec:conclusions}

We have introduced a novel algorithm that decomposes streak camera images into distributions of photoelectrons. This algorithm relies on an image deconvolution technique that uses premeasured photoelectron profiles to create deconvolution kernels. The image clusters formed by single photoelectrons are characterized in terms of widths in pixels and amplitude.  The cluster widths are found to be largely independent of the sweep speed of the streak camera. Performing the image decomposition into a distribution of photoelectrons in space and time enables the application of traditional statistical analyses, including counting statistics, on the resultant image features and to report individual uncertainty measurement for each feature in the image.

We tested our algorithm and its uncertainty prediction for timing measurements using a dataset where laser pulses, separated by a fixed time interval, were available as well as with AWAKE data sets. The dataset with multiple laser pulse features  was produced by introducing a glass pane in the path of a laser pulse, generating several image features due to reflections in the glass.  The time differences between the different reflections were then analyzed.  For the AWAKE data set, the image from a single laser pulse was spatially split and the timing measurements of the two parts compared. The time resolution was found to be $\sigma_t = 7.50/{\sqrt{n_{\gamma}}}$~ps, for $210$~ps sweep time and $\sigma_t = 2.61/{\sqrt{n_{\gamma}}}$~ps, for $73$~ps sweep time. This simple formula was verified for $400 \leq n_{\gamma} \leq 3000$.

\acknowledgments

We would like to express our gratitude to the AWAKE Collaboration for providing the glass pane reflection dataset and the proton beam dataset. We expressly thank Fabian Batsch for his help in understanding the specifics of the reflection dataset which he set up and generated.

\bibliographystyle{JHEP}
\bibliography{biblio}

\providecommand{\href}[2]{#2}\begingroup\raggedright\begin{thebibliography}{10}

\bibitem{AWAKE:2022kmf}
{L. Verra et al. (AWAKE Collaboration)}, \emph{{Controlled Growth of the Self-Modulation of a Relativistic Proton Bunch in Plasma}}, \href{https://doi.org/10.1103/PhysRevLett.129.024802}{\emph{Phys. Rev. Lett.} {\bfseries 129} (2022) 024802} [\href{https://arxiv.org/abs/2203.13752}{{\ttfamily 2203.13752}}].

\bibitem{Bachmann:2020eoj}
{A.-M. Bachmann and P. Muggli}, \emph{{Beam Diagnostics in the Advanced Plasma Wakefield Experiment AWAKE}}, \href{https://doi.org/10.18429/JACoW-IBIC2020-MOAO01}{\emph{IBIC} (2020) MOAO01}.

\bibitem{Hogan_2003}
M.J.~Hogan, C.E.~Clayton, C.~Huang, P.~Muggli, S.~Wang, B.E.~Blue et~al., \emph{Ultrarelativistic-positron-beam transport through meter-scale plasmas}, \href{https://doi.org/10.1103/PhysRevLett.90.205002}{\emph{Phys. Rev. Lett.} {\bfseries 90} (2003) 205002}.

\bibitem{Rieger_2017}
K.~Rieger, A.~Caldwell, O.~Reimann and P.~Muggli, \emph{{GHz} modulation detection using a streak camera: {Suitability} of streak cameras in the {AWAKE} experiment}, \href{https://doi.org/10.1063/1.4975380}{\emph{Rev. Sci. Instrum.} {\bfseries 88} (2017) 025110}.

\bibitem{Bachmann_2020}
A.-M.~Bachmann and P.~Muggli, \emph{Determination of the {Charge} per {Micro}-{Bunch} of a {Self}-{Modulated} {Proton} {Bunch} using a {Streak} {Camera}}, \href{https://doi.org/10.1088/1742-6596/1596/1/012005}{\emph{J. Phys. Conf. Ser.} {\bfseries 1596} (2020) 012005}.

\bibitem{Rossa_1992}
E.~Rossa, C.~Bovet, L.~Disdier, F.~Madeline and J.-J.~Savioz, \emph{Real Time Measurement of Bunch Instabilities in LEP in three Dimensions using a Streak Camera}, Conference Proceedings, Editions Frontieres, Paris, France (1992).

\bibitem{Suntao_2018}
S.~Wang and D.~Rubin, \emph{Measurement of the beam yz crabbing tilt due to wakefields using streak camera at {CESR}}, \href{https://doi.org/10.1088/1742-6596/1067/7/072010}{\emph{J. Phys. Conf. Ser.} {\bfseries 1067} (2018) 072010}.

\bibitem{Oz_2016}
E.~Öz, F.~Batsch and P.~Muggli, \emph{An accurate rb density measurement method for a plasma wakefield accelerator experiment using a novel rb reservoir}, \href{https://doi.org/10.1016/j.nima.2016.02.005}{\emph{Nucl. Instrum. Methods Phys. Res. A: Accel. Spectrom. Detect. Assoc. Equip.} {\bfseries 829} (2016) 321}.

\bibitem{Caldwell:2015rkk}
{A. Caldwell et al. (AWAKE Collaboration)}, \emph{{Path to AWAKE: Evolution of the concept}}, \href{https://doi.org/10.1016/j.nima.2015.12.050}{\emph{Nucl. Instrum. Meth. A} {\bfseries 829} (2016) 3} [\href{https://arxiv.org/abs/1511.09032}{{\ttfamily 1511.09032}}].

\bibitem{AWAKE:2015taz}
{E. Gschwendtner et al. (AWAKE Collaboration)}, \emph{{AWAKE, The Advanced Proton Driven Plasma Wakefield Acceleration Experiment at CERN}}, \href{https://doi.org/10.1016/j.nima.2016.02.026}{\emph{Nucl. Instrum. Meth. A} {\bfseries 829} (2016) 76} [\href{https://arxiv.org/abs/1512.05498}{{\ttfamily 1512.05498}}].

\bibitem{AWAKE:2022aeo}
{E. Gschwendtner et al. (AWAKE Collaboration)}, \emph{{The AWAKE Run 2 Programme and Beyond}}, \href{https://doi.org/10.3390/sym14081680}{\emph{Symmetry} {\bfseries 14} (2022) 1680} [\href{https://arxiv.org/abs/2206.06040}{{\ttfamily 2206.06040}}].

\bibitem{Batsch:2019vsw}
F.~Batsch, \emph{{Setup and Characteristics of a Timing Reference Signal with sub-ps Accuracy for AWAKE}}, \href{https://doi.org/10.1088/1742-6596/1596/1/012006}{\emph{J. Phys. Conf. Ser.} {\bfseries 1596} (2020) 012006} [\href{https://arxiv.org/abs/1911.12201}{{\ttfamily 1911.12201}}].

\bibitem{AWAKE:2018wrd}
{E. Adli et al. (AWAKE Collaboration)}, \emph{{Experimental observation of proton bunch modulation in a plasma, at varying plasma densities}}, \href{https://doi.org/10.1103/PhysRevLett.122.054802}{\emph{Phys. Rev. Lett.} {\bfseries 122} (2019) 054802} [\href{https://arxiv.org/abs/1809.04478}{{\ttfamily 1809.04478}}].

\bibitem{AWAKE:2020stp}
{F. Batsch et al. (AWAKE Collaboration)}, \emph{{Transition between Instability and Seeded Self-Modulation of a Relativistic Particle Bunch in Plasma}}, \href{https://doi.org/10.1103/PhysRevLett.126.164802}{\emph{Phys. Rev. Lett.} {\bfseries 126} (2021) 164802} [\href{https://arxiv.org/abs/2012.09676}{{\ttfamily 2012.09676}}].

\bibitem{Nechaeva_2023hosing}
{T. Nechaeva et al. (AWAKE Collaboration)}, \emph{Hosing of a long relativistic particle bunch in plasma},  \href{https://arxiv.org/abs/2309.03785}{{\ttfamily 2309.03785}}.

\bibitem{Fedosseev:2016ccm}
V.~Fedosseev et~al., \emph{{Integration of a Terawatt Laser at the CERN SPS Beam for the AWAKE Experiment on Proton-Driven Plasma Wake Acceleration}},  in \emph{{7th International Particle Accelerator Conference}}, p.~WEPMY020, 2016, \href{https://doi.org/10.18429/JACoW-IPAC2016-WEPMY020}{DOI}.

\bibitem{Hamamatsu:2022}
P.K.~Hamamatsu, \emph{Universal streak camera C10910-05 Catalog}, Hamamatsu, Photonics K.k (2022).

\bibitem{Wing:2643452}
{M. Wing et al. (AWAKE Collaboration)}, \emph{{AWAKE Status Report}},  Tech. Rep. \href{https://cds.cern.ch/record/2643452}{CERN-SPSC-2018-030, SPSC-SR-240}, CERN, Geneva (2018).

\bibitem{Gonzalez_2008}
R.C.~Gonzalez and R.E.~Woods, \emph{Digital image processing}, Prentice Hall, Upper Saddle River, N.J. (2008).

\bibitem{Rosenfeld:1966}
A.~Rosenfeld and J.L.~Pfaltz, \emph{Sequential operations in digital picture processing}, \href{https://doi.org/10.1145/321356.321357}{\emph{J. ACM} {\bfseries 13} (1966) 471}.

\end{thebibliography}\endgroup

\end{document}